\documentclass[12pt]{article}

\usepackage{epsf}
\usepackage[dvips]{graphicx}

\begin{document}

\begin{center}
{\bfseries MULTIPARTICLE PRODUCTION AT HIGH MULTIPLICITIES}

\vskip 5mm

E.S. Kokoulina$^{1,2 \dag}$, A.Y. Kutov$^{1,3}$: for {\bfseries
SVD-2 Collaboration}

\vskip 5mm

{$^1$ {\it LPS, GSTU, Belarus}, $^2$ {\it LPP, JINR, Russia } $^3$
{DM UrD,RAS,Russia}
\\
$\dag$ {\it E-mail: kokoulin@sunse.jinr.ru }}
\end{center}

\vskip 5mm
%
%

\abstract{Theoretical and experimental studies of high multiplicity
events are analyzed. Some interesting phenomena can be revealed at
high multiplicities. Preliminary results of project $\it
{Thermalization}$ are reported.}

\section{Introduction}
The multiparticle production (MP) study at high energies is one of
the actual topics of high energy physics. The different theoretical
approaches and the experimental programs are developed. The
Quark-Gluon Matter search is the complicated problem of hadron and
nucleus interactions \cite{LHC}. We consider that our MP study at
lower energies may be useful. The purpose of the $\it
{"Thermalization"}$ experiment \cite{THERM} is to investigate the
collective behavior of MP particles in proton and proton-nucleus
interactions
\begin{equation}
\label{1} p+p(A) \rightarrow n_{\pi} \pi +2N
\end{equation}
at the proton energy $E_{lab}=70$ GeV. We use modernized setup SVD-2
- Spectrometer with Vertex Detector (SVD). It was created to study
of production and decay of charm particles, but had the basic
components necessary for performance of the physical program of the
$\it {Thermalization}$ project.\

At present multiplicity distributions (MD) at this energy is
measured up to the number of charged particles $n_{ch}=18$
(\cite{Bab}-\cite{Bog}). In the region of high multiplicity (HM)
$n_{ch}>20$ we expect \cite{Prop}: formation of high density
thermalized hadronic system, transition to pion condensate or cold
QGP, increase of partial cross section $\sigma (n)$ is expected in
comparison with commonly accepted extrapolation, enhanced rate of
direct soft photons. We will be continue to search for new
phenomena: Bose-Einstein condensate (BEC), events with ring topology
(Cherenkov gluon radiation). The available MP models and MC codes
(PYTHIA) are distinguished considerably at the HM region. We also
research hadronization mechanism and connected questions \cite{HM}.

The review is organized as follows: section 2 presents a description
of setup SVD-2, section 3 gives alignment results, section 4 informs
about of new phenomena searching and our preliminary data of 2002
run. We summarize in section 5.
\section{Experimental setup}
\subsection{Setup schematic}

The layout of the SVD installation at $U-70$ accelerator is shown on
Figure 1.
The basic requirements to the equipment consisted in the following:\\
$\ast $ The study is carried out on the extracted beam of protons
with energy 70 GeV and intensity $\sim 10^7 $ in a cycle of the
accelerator.\\
$\ast $ The liquid hydrogen target is used.\\
$\ast $ Installation is capable to detect of events with HM of
charged particles and $\gamma $ quanta. Multiplicity of photons
makes up to $\leq 100$. The lower energy threshold of the photon
registration is ~50 MeV.\\
$\ast $ The HM trigger system is capable to select rare events with
multiplicity $n_{\pi }=20\div 30$. The suppression factor of
events with low multiplicity $n_{\pi }<20$ is $10^4$.\\
$\ast $ The magnetic spectrometer has the momentum resolution
$\delta p/p\approx1.5\%$ in the interval $p=0.3\div 5.0$ GeV/c.\ At
the beginning the experiment and subsequent data analysis the
generator was developed. It is based on the assumption that in the
HM region the particles in c.m.s. should have isotropic angular
distribution and their energy distribution is Maxwell or
Bose-Einstein type \cite{Prop}.
\begin{figure}[h!]
     \leavevmode
\centering
\includegraphics[width=3.3in, height=2.3in, angle=0]{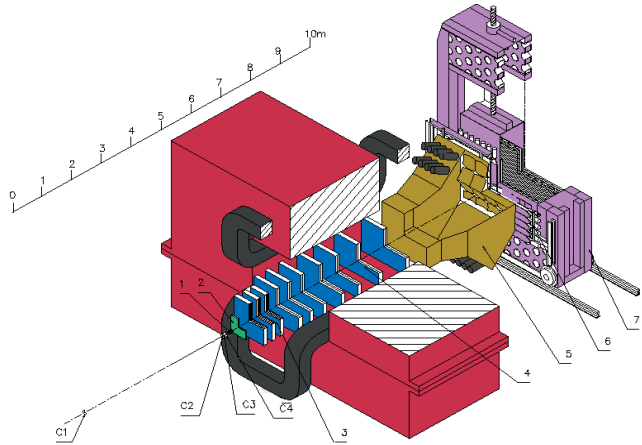}
\caption{Schematic of the SVD installation at U-70. }
\end{figure}

\subsection{Liquid hydrogen target}
For a target accommodation in the design of installation there is a
space along a beam only 7 cm. Design and manufacture of liquid
hydrogen target is under a a complete JINR responsibility. The
target has 7 cm thick and 3.5 cm in diameter vassal of liquid
hydrogen. Thermostat is equipped with a thin (200 $\mu m$) lavsan
windows to suppress background scattering. Successful tests of a
whole target system had indicated to advanced reduction of helium
consumption in which resulting factor is expected in order of 1.5.
\subsection{Straw tube chambers}
Straw tube chamber system is a new addition of SVD setup. This
detector has been designed in the department of V.~Peshehonov from
LPP of JINR . It implements front end boards with preamplifiers
produced in Minsk (NC PHEP BSU) and TDC modules produced in Protvino
(IHEP) allowing to detect several pulses, consequently coming from
the anode on each trigger signal. Typical plane dimension is 70 x 70
$cm^2$. The total of channels is about 2400.
\subsection{HM trigger}
Our experiment owes to carry out at suppression of low multiplicity
events by a trigger. It is urgent request for it. For this purpose
the scintillation hodoscope or HM trigger was designed and
manufactured. It suppresses interactions with track multiplicity
below 20. Beyond this it is as so thin as not distorts an angular
and momentum resolution of the setup to any kind fake signal. The
scintillator counter array may operate at higher counting rate and
more resistant to many kinds of noise.
\subsection{Vertex detector}
The vertex detector (VD) is  necessary constituent of SVD setup
because it allows to vertex position identify. Vertex front-end uses
a integrated circuit called GASSIPLEX. As the GASSIPLEX is
16-channel design, only 1280 channels of detector may be placed on
one board. For 50 $\mu m$ pitch detector the largest sensitive area
dimension is 64 $mm$. To overcome this restriction the Collaboration
had taken the decision to use integrated 128-channel circuits
VIKING. JINR provides important technical support in this
development. Now we had purchased a requisite consignment of these
circuits and are installing in VD.
\subsection{Magnetic spectrometer, Gamma-detector}
The magnet MC-7A having length on the beam 3 $m$ is used in
spectrometer. Magnet field in the center is equal to 1.1 $T$ at a
current 4000 $A$. The detection system of the spectrometer includes
18 planes of proportional chambers (PC). The data analysis give the
following characteristics of the spectrometer: average PC efficiency
is 80$\%$; coordinate accuracy on the reconstructed tracks is 1
$mm$; the momentum resolution on beam tracks (p=70 GeV/c) is 3 $\%$;
the momentum resolution on the secondary tracks is $\sim $1 $\%$.
Magnetic spectrometer electronics allows to register up to 1.5
thousand events per 1 accelerator cycle. Some of PC had been
repaired, anode wires in beam region are covered with insulator to
make them insensitive to beam particles. This modification improves
efficiency of central part of chamber at high beam intensity $10^7$
required for $\it {Thermalization}$ project.

The gamma-detector consists of 1536 full absorption Cherenkov
counters. Radiators from a lead glass have the size $38 \times 38
\times 505$ $mm^3$ and are connected with PMT-84-3. Total fiducial
area of the detector is $1.8 \times 1.2$ $m^2$. The energy
resolution on 15 $GeV$ electrons is $12 \% $. Accuracy of the
$\gamma $ quantum coordinate reconstruction is $\sim 2$ $mm$. At run
2007 the gamma-detector calibration was carried out and gamma-
quantum events were recorded.
\section{Alignment}
The importance task of any experiment is to provide reconstruction
of charged particle tracks. Spatial characteristics and geometric
position of detector modules can be differ from its design values.
Possible reasons of detector misalignments are the limited accuracy
of initial hardware, inaccuracies in placing of detectors and their
internal dimensions. The alignment procedure intends to compensate
such misalignment by a mathematical way. We use for alignment
procedure more robust, efficient and high precision method based on
the Linear Least Squares (LLS) \cite{Blob}.

At 2006 technical run we had obtained data on hydrogen target. We
had picked out some events with good identification of 787 (single)
space tracks on hits in vertex detector and carried out alignment.
Histograms of $\chi ^2/n_{df}$ for local fits before and after
alignment procedure are in Figure 2.
\begin{figure}[h!]
\leavevmode \centering
\includegraphics[width=1.6in, height=1.5in, angle=0]{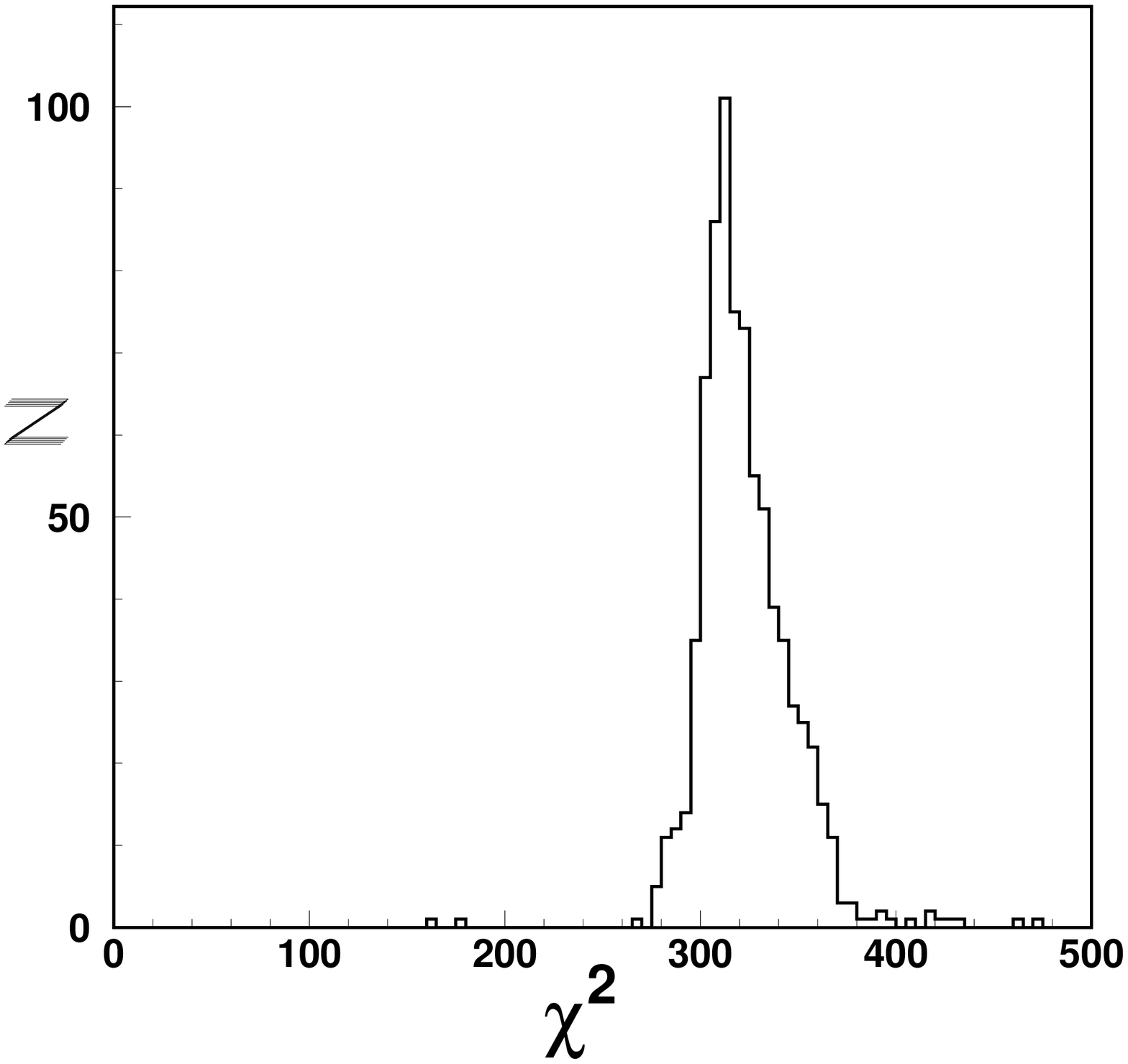}
\includegraphics[width=1.6in, height=1.5in, angle=0]{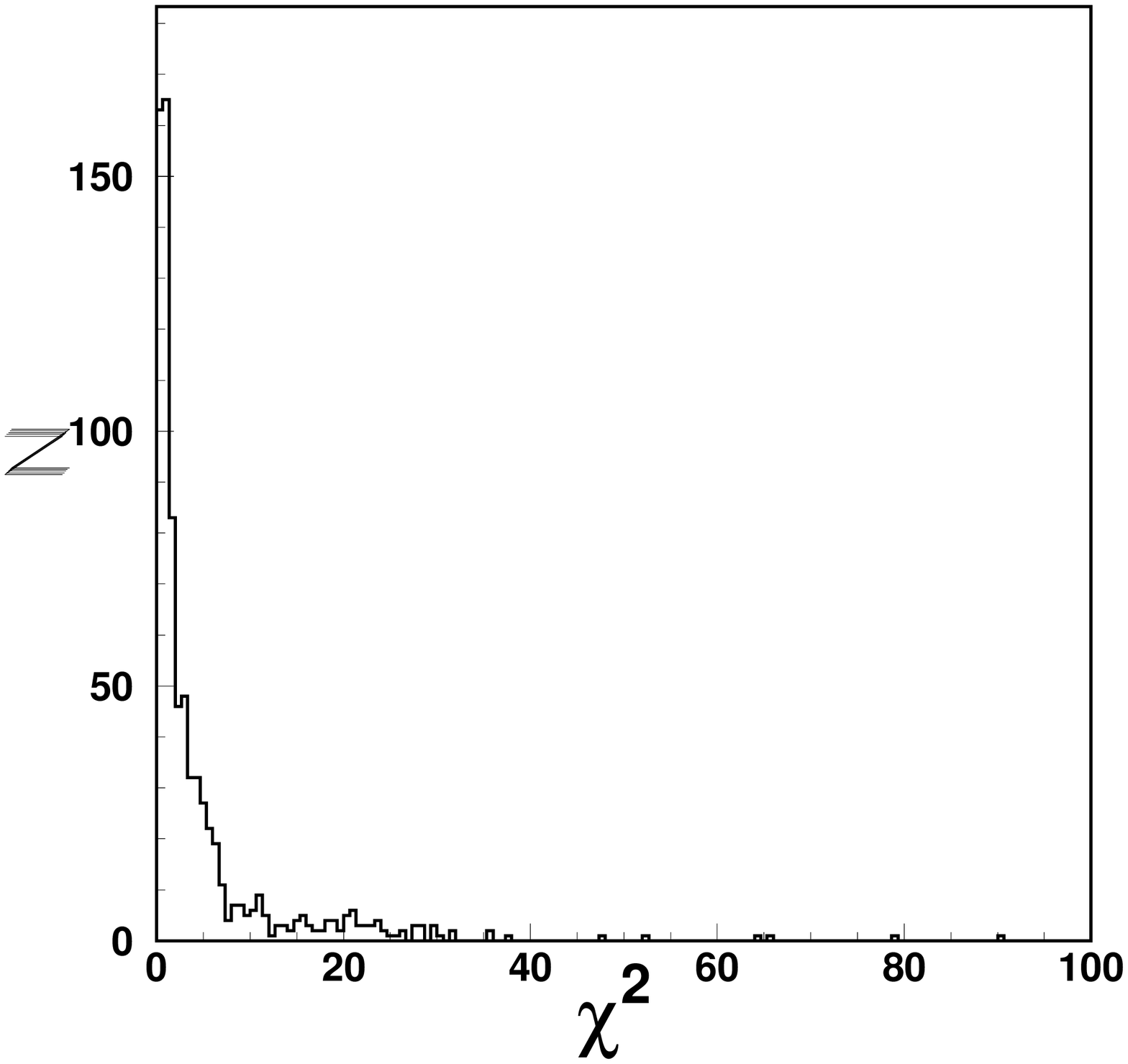}
\caption{$\chi ^2/n_{df}$ for tracks: (left) before and (right)
after alignment.}
\end{figure}
At present it is continued data processing and high multiplicity
event searching. One of such events is shown on Figure 3.
\begin{figure}[h!]
     \leavevmode \centering
\includegraphics[width=3.4in, height=4.1 in, angle=270]{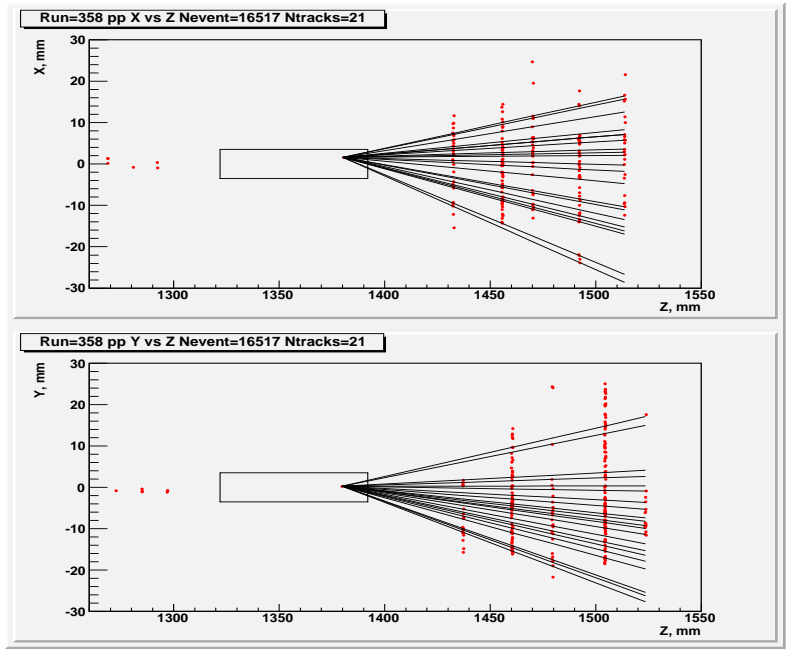}
\caption{Event with multiplicity 21. }
\end{figure}
Preliminary multiplicity distribution of charged particles was
obtained based on VERTEX detector data. It is shown on Figure 4.
\begin{figure}[h!]
     \leavevmode \centering
\includegraphics[width=4.1in, height=3.5in, angle=0]{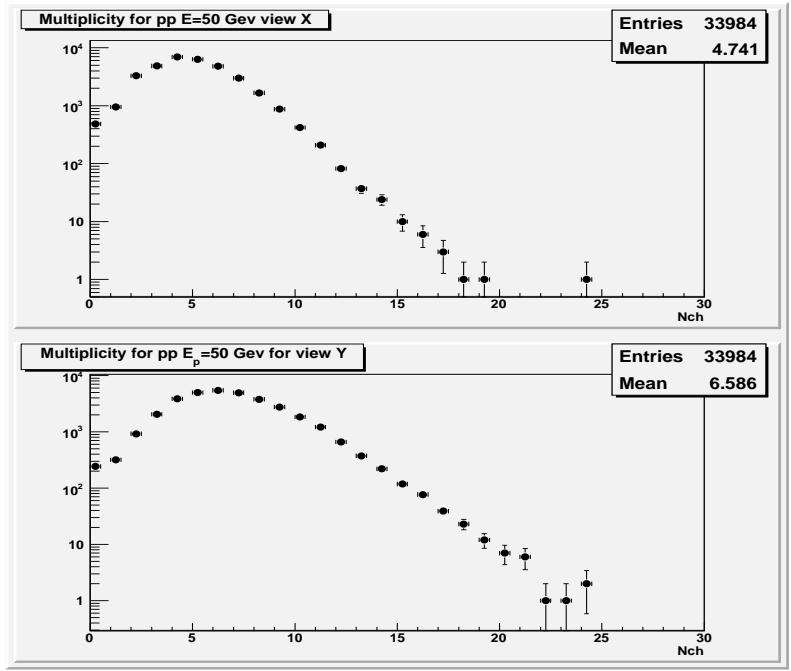}
\caption{Preliminary MD in pp at HM region. }
\end{figure}

\section{Search for new phenomena}
The HM region study is important, because MP models and Monte-Carlo
generators are differed at high multiplicity ($n > \overline n(s)$)
very considerably. There are the theoretical predictions about
manifestation such phenomena as Cherenkov-like (gluon) radiation
\cite{Drem}, Bose-Einstein condensation (BEC) of pions \cite{BEC,
Gor}, excess of soft photon rate \cite{CHL} and other collective
phenomena. We like to reveal their in our findings.

For multiparticle dynamics insight and the MD description in hadron
interactions we had proposed the Gluon Dominance model (GDM)
\cite{GDM}. In the framework of this model we research quark-gluon
matter and hadronization stage detail by using MD of the charged and
neutral particles and their moments \cite{GDM1}. GDM bases on the
essentials of QCD and phenomenological scheme of hadronization. Our
model studies had shown: valent quarks of initial protons are
staying in leading particles (from 70 to 800 GeV/c and higher). MP
is realized by gluons. We called them active ones.

Some of active gluons ($\sim 50\%$) are staying inside quark-gluon
system and do not fragment to hadrons. New formed hadrons catching
up them, are excited and throw down excess of energy by soft photons
(SP). We use the black body emission spectrum at the assumption that
quark-gluon system or excited new formed hadrons set in almost
equilibrium state during a short period. This assumption permits to
estimate the line size of the SP emission region \cite{Kur}. It is
known that in this region hadronization is occurred.

Our model confirms the recombination mechanism of MP. We had
obtained limitations on the number charged, neutral and total
multiplicities in $pp$ interactions at 70 GeV/c and higher. In
project Thermalization we plan to verify these. There are many of
experimental and theoretical results, which evidence of cluster
nature of MP by significant short-range multiplicity correlations
\cite{Rol}, the observed scaling of the dynamical fluctuations of
mean transverse-momentum \cite{Flo}.

In GDM the evaporation of gluon sources may be realized by single
gluons and also groups consisted from two or more fission gluons.
The superposition of them explains the shoulder structure of MD at
ISR and higher energies \cite{GDM}. Our approach gives the possible
interpretation of soft and semi-hard components \cite{GIO4}.

We modified GDM by including of the intermediate quark topologies to
explain the experimental differences between $p\overline p$ and $pp$
inelastic topological cross sections and second correlation moment
behavior at few GeV/c \cite{Rush}. The high multiplicity in this
process originates from "4" or "6"-topologies. Our scheme of
hadronization describes well MD for hadron interactions at 70 GeV
and higher and could be use to study the central nuclear collisions
at low and high energies.
\begin{figure}[h!]
\leavevmode \centering
\includegraphics[width=5.3 in, height=2.2 in, angle=0]{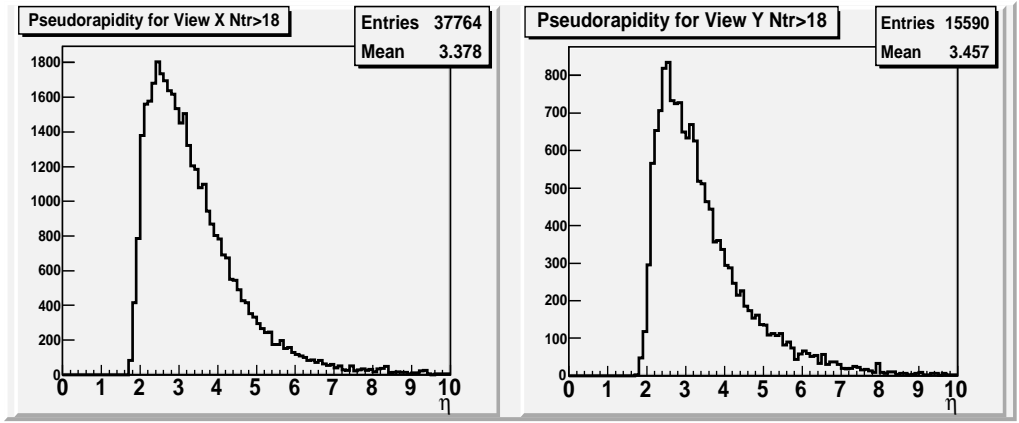}
\caption{The pseudorapidity spectra in $pPb$ at $n>18$.}
\end{figure}

The Cherenkov type radiation can be emitted in the projectile and
target particles. This leads to two peaks of dense groups of
particles (spikes) distribution in rapidity phase-space. At the same
time the particle distribution at the azimuthal angle is uniform.
Study of the spike center distribution \cite{EGS} in central C-Cu
collisions at 4.5~GeV/c/A (all charged particles) and Mg-Mg
collisions at 4.3~GeV/c/A (only negative charged particles) were
found to be in agreement with the hypothesis of mesonic Cherenkov
radiation. For this goal it was used transformation of
pseudorapidity spectra from $\eta$ variable to $\tilde\eta$ with the
uniform spectrum. In each case the distance between peaks for these
experiments is in agreement with Cherenkov radiation hypothesis, the
charged-dependence was absent.

The ring-like substructures of secondary in $^{208}Pb$ at $158$ A
GeV/c and $^{197}Au$ at 11.6 GeV/c induced interactions with Ag(Br)
nuclei in emulsion detector were investigated \cite{Vok}. The good
agreement was obtained with idea of Cherenkov radiation.

It must be emphasized that such events are rare, and represent at
about 1\% of full statistics. Therefore high luminosity and high
multiplicity trigger of SVD setup agrees to collect enough
statistics to study this phenomenon. The preliminary indications to
the manifestation of the ring events are in Figure 5. This
pseudorapidity spectra for $pPb$-interactions at high multiplicity
($n>18$) shows up such behavior.

As it was mentioned the Bose-Einstein condensation is very
interesting phenomenon. The considerable efforts are necessary to
confirm it experimentally. At HM events the plentiful number of
pions (charged and neutral) are produced. All of them are bosons.
When the multiplicity increase moments of them are approaching to
zero. In the case of relativistic ideal Bose gas the pion number
fluctuations may give a clear signal of approaching the BEC point
\cite{Gor}. When the temperature $T$ decreases, the chemical
potential increases and becomes equal to $\mu _{\pi}$=$m_{\pi}$ at
BEC temperature $T=T_C$. At this point the total number of particles
takes up the lowest energy state.

M.I.~Gorenstein and V.V.~Begun had viewed the case of HM events in
$p+p$ interactions with beam energy of $70$ GeV \cite{Gor}. The
volume of pion system was estimated as, $V=E/\varepsilon (T, \mu
_{\pi })$, and the number of pions was determined as, $N_{\pi }=V
\rho_{\pi } (T, \mu _{\pi })$. In the vicinity of the BEC point they
revealed an abrupt and anomalous increase of the scaled variance of
neutral and charged pion number fluctuations. Our experiment permits
to experimental test of this phenomena. We are expected to take a
lot of high multiplicity event statistics with reconstructed  by
gamma quantum neutral mesons and study scaled variance of neutral
and charged pion number fluctuations,

\section{Summary}
We are continueing our work to making of program packets for data
processing and new phenomena study at HM region.

\section{Acknowledges}
Author E.K. is glad to thank the NPQCD-2007 Org.Committee for
partial financial support and warm working atmosphere created.

These researches implemented into framework of {\it project
"Thermalization" is partially supported by RFBR grant
$06-02-81010-Bel\_ a$}.

\end{document}